# A conservative scheme for electromagnetic simulation of magnetized plasmas with kinetic electrons

## J. Bao[1,2], Z. Lin[1,*] and Z. X. Lu[2, 3]

[1] Department of Physics and Astronomy, University of California, Irvine, California 92697, USA

[2] Fusion Simulation Center, Peking University, Beijing 100871, People's Republic of China

[3] Max Planck Institut für Plasmaphysik, Garching, Germany

[*] Corresponding author: E-mail: zhihongl@uci.edu

A conservative scheme has been formulated and verified for gyrokinetic particle simulations of electromagnetic waves and instabilities in magnetized plasmas. An electron continuity equation derived from drift kinetic equation is used to time advance electron density perturbation by using the perturbed mechanical flow calculated from the parallel vector potential, and the parallel vector potential is solved by using the perturbed canonical flow from the perturbed distribution function. In gyrokinetic particle simulations using this new scheme, shear Alfvén wave dispersion relation in shearless slab and continuum damping in sheared cylinder have been recovered. The new scheme overcomes the stringent requirement in conventional perturbative simulation method that perpendicular grid size needs to be as small as electron collisionless skin depth even for the long wavelength Alfvén waves. The new scheme also avoids the problem in conventional method that an unphysically large parallel electric field arises due to the inconsistency between electrostatic potential calculated from the perturbed density and vector potential calculated from the perturbed canonical flow. Finally, the gyrokinetic particle simulations of the Alfvén waves in sheared cylinder have superior numerical properties compared with the fluid simulations, which suffer from numerical difficulties associated with singular mode structures.

## I. Introduction

The kinetic effects of electrons are important to long wavelength magnetohydrodynamic(MHD) instabilities and short wavelength drift-Alfvénic instabilities responsible for turbulence transport in magnetized plasmas, since the non-adiabatic electron can interact with, modify and drive the low frequency instabilities[1, 2]. Electromagnetic simulation with kinetic electrons encounters great computational challenges, especially for the long wavelength modes in high $\beta$ plasmas ( $\beta$ is the ratio between kinetic and magnetic pressures)[3, 4], which have been widely studied during the past two decades. In particular, the electromagnetic gyrokinetic particle simulation directly solving the electron drift kinetic equation can not address efficiently the long wavelength modes, and the numerical performance becomes worse in high $\beta$ plasmas. Early studies suggest that the perpendicular grid size should resolve electron skin depth for conventional perturbative ($\delta f$) simulations of shear Alfvén waves[4], which represents a huge computational cost for the global simulation. The origin of this problem is that the large two adiabatic terms on both sides of the parallel Ampere's law using canonical momentum $p_{||}$ formulation can not cancel with each other exactly when the grid size is larger than the electron skin depth in conventional $\delta f$ simulation[5, 6], which is sometimes referred as "cancellation problem" that affects the accuracy of the small non-adiabatic part [7, 8]. Although many methods have been developed to overcome or avoid the "cancellation problem" [5-15], electromagnetic gyrokinetic particle simulation with kinetic electrons is still computationally challenging.

In this work, we have formulated and verified a new conservative scheme using both electron continuity equation and drift kinetic equation for gyrokinetic simulation of electromagnetic modes including long wavelength shear Alfvén waves. To demonstrate the successful application of conservative scheme, we focus on the $p_{||}$ formulation which suffers severe numerical problem in early numerical studies [3, 4]. In conventional $\delta f$ algorithm, it is found that the electron perturbed density and canonical flow measured from kinetic markers do not satisfy the electron continuity equation. Consequently, the electrostatic potential calculated from the density and the parallel vector potential calculated from the canonical flow are not consistent with each other, which results in an unphysically large parallel electric field. In the new conservative scheme, we use the electron continuity equation to time advance the electron density perturbation using the perturbed mechanical flow calculated from the parallel vector potential, and the parallel vector potential is solved by using the perturbed canonical flow from the kinetic markers. Using this new scheme in gyrokinetic particle simulation, we have verified the dispersion relation of shear Alfvén wave in the shearless magnetic field, and Alfvén continuum damping and kinetic Alfvén wave propagation in the sheared magnetic field. This new conservative scheme has no restriction of the perpendicular grid size to resolve the electron skin depth for the long wavelength MHD wave simulation, and only requires a very small number of kinetic markers (20 markers per cell) to achieve sufficient accuracy in linear simulation. In addition to the simulations of the low frequency drift-Alfvénic modes described in the paper, the new conservative scheme has also been successfully applied to the simulations of radio frequency (RF) waves such as lower hybrid wave in tokamak plasmas[16, 17]. Both the simulation and theory have shown that an unphysically large $\delta E_{||}$ is produced due to the inconsistency of the perturbed density and canonical flow in the simulation of the long wavelength shear Alfvén wave in conventional $\delta f$ scheme, which leads to the well-known numerical difficulties. In contrast, conservative scheme guarantees the correctness of $\delta E_{||}$ by enforcing the consistency between the perturbed density and canonical flow, and thus enforcing the consistency between the electrostatic potential and the parallel vector potential. Finally, by comparing the gyrokinetic particle simulation and fluid simulation of the kinetic Alfvén wave propagation in the sheared plasmas, we found that the fluid simulation suffers the artificial singular mode structure problem due to the model limitation, while gyrokinetic particle simulation physically avoids the artificial singularity by inducing the kinetic effects, i.e., electron Landau damping, which shows the superiority of our gyrokinetic particle simulation model.

The paper is organized as follows: The physics model of conservative scheme for electromagnetic simulations with drift kinetic electrons is introduced in Sec. II. The simulation results of shear Alfvén waves in both the uniform and sheared magnetic field are shown in Sec. III. In Sec. IV, the effects of particle noise in the PIC simulations of long wavelength shear Alfvén waves are studied theoretically for conventional perturbative scheme and conservative scheme. Sec. V is the conclusion.

## II.    Conservative scheme

Electromagnetic simulation with kinetic electrons suffers severely numerical instabilities when $\beta_e \gg m_e/m_i$ and $k_\perp^2 d_e^2 \ll 1$ ($k_\perp$ is the perpendicular wave vector, $d_e$ is the electron skin depth) in earlier gyrokinetic particle simulations[3, 4]. In this work, we are able to avoid these numerical instabilities by using an electron continuity equation to ensure the density conservation in the simulation, which is referred as the conservative scheme. The ordering adopted in the paper is:

$$\frac{\omega}{\Omega_i} \sim \frac{\delta f}{f} \sim \frac{e\phi}{T} \sim \frac{\delta B}{B_0} \sim \frac{k_{||}}{k_\perp} \sim \frac{1}{k_\perp L} \sim O(\varepsilon) << 1,$$

where $\omega$ and $\Omega_i = eB_0/cm_i$ are physical mode frequency and ion cyclotron frequency, $\delta f$ and $f$ are perturbed and total particle distributions, $\phi$ and $\delta B$ are perturbed electrostatic potential and perturbed magnetic field, and $k_\perp$ and $k_{||}$ are the perpendicular and parallel wave vectors. $L$ is plasma equilibrium scale length $L \sim (L_n = \nabla \ln n, L_T = \nabla \ln T, L_B = \nabla \ln B_0)$ ($T$ is plasma temperature, $n$ is plasma density, and $B_0$ is equilibrium magnetic field).

The electron drift kinetic equation uses guiding center position $\mathbf{R}$, parallel canonical momentum $p_{||} = m_e v_{||} + q_e \delta A_{||}/c$ ($m_e$ and $q_e$ denote electron mass and charge, respectively) and magnetic momentum $\mu$ as independent variables in five dimensional phase space[18, 19]:

$$\frac{\partial}{\partial t}\left(f_e B_{||}^*\right) + \nabla \cdot \left(\dot{\mathbf{R}} f_e B_{||}^*\right) + \frac{\partial}{\partial p_{||}}\left(\dot{p}_{||} f_e B_{||}^*\right) = 0, \tag{1}$$

$$\dot{\mathbf{R}} = \frac{1}{m_e}\left(p_{||} - \frac{q_e}{c}\delta A_{||}\right)\frac{\mathbf{B_0^*}}{B_{||}^*} + \frac{c\mathbf{b_0}}{q_e B_{||}^*}\times\left(\mu\nabla B_0 + q_e\nabla\Psi_l + q_e\nabla\Psi_{nl}\right),$$ (2)

$$\dot{p}_{||} = -\frac{\mathbf{B_0^*}}{B_{||}^*}\cdot\left(\mu\nabla B_0 + q_e\nabla\Psi_l + q_e\nabla\Psi_{nl}\right),$$ (3)

where $f_e = f_e\left(\mathbf{R}, p_{||}, \mu, t\right)$ is the electron guiding center distribution, $\mathbf{B_0} = B_0\mathbf{b_0}$ is the equilibrium magnetic field, $\mathbf{B_0^*} = \mathbf{B_0} + \frac{c}{q_e}p_{||}\nabla\times\mathbf{b_0}$, $B_{||}^* = \mathbf{b_0}\cdot\mathbf{B_0^*}$, $\phi$ is the electrostatic potential, $\delta A_{||}$ is the vector potential, and $\Psi_l = \phi - \frac{p_{||}\delta A_{||}}{m_e c}$ and $\Psi_{nl} = \frac{q_e\delta A_{||}^2}{2m_e c^2}$ are the linear and nonlinear parts of the generalized potential, respectively. Eqs. (1)-(3) are also referred as $p_{||}$ formulation in a conservative form and satisfy the Liouville's theorem [19].

In order to minimize the particle noise, the perturbative ($\delta f$) method [20, 21] is widely used in particle-in-cell simulations. In perturbative ($\delta f$) method, the distribution function $f_e$ is decomposed into equilibrium and perturbed parts as $f_e = f_{e0} + \delta f_e$. In the lowest order equilibrium part, Eq. (1) reduces to:

$$L_0 f_{e0} = 0.$$ (4)

where $L_0 = \frac{\partial}{\partial t} + \left(\frac{p_{||}}{m_e B_{||}^*}\mathbf{B_0^*} + \frac{c\mu}{q_e B_{||}^*}\mathbf{b_0}\times\nabla B_0\right)\cdot\nabla - \frac{\mu}{B_{||}^*}\mathbf{B_0^*}\cdot\nabla B_0\frac{\partial}{\partial p_{||}}$ is the equilibrium propagator. The equilibrium solution $f_{e0}$ can be approximated as a Maxwellian:

$$f_{e0} = n_{e0}\left(\frac{m_e}{2\pi T_{e0}}\right)^{3/2}\exp\left(-\frac{p_{||}^2/m_e + 2\mu B}{2T_{e0}}\right)$$ when the neoclassical effects are not important.

Subtracting Eq. (1) by Eq. (4), the equation for the perturbed distribution $\delta f_e$ is

$$L\delta f_e = -\left(\delta L_1 + \delta L_2\right)f_{e0},$$ (5)

where $L = \frac{\partial}{\partial t} + \dot{\mathbf{R}}\cdot\nabla + \dot{p}_{||}\frac{\partial}{\partial p_{||}}$ is the total propagator, and the perturbed propagator $\delta L = \delta L^L + \delta L^{NL}$ consists of linear and nonlinear parts:

$$\delta L^L = \left( -\frac{q_e \delta A_{||}}{cm_e} \frac{\mathbf{B_0^*}}{B_{||}^*} + \frac{c\mathbf{b_0} \times \nabla \Psi_l}{B_{||}^*} \right) \cdot \nabla - q_e \frac{\mathbf{B_0^*}}{B_{||}^*} \cdot \nabla \Psi_l \frac{\partial}{\partial p_{||}},$$

$$\delta L^{NL} = \frac{c\mathbf{b_0}}{q_e B_{||}^*} \times q_e \nabla \Psi_{nl} \cdot \nabla - q_e \frac{\mathbf{B_0^*}}{B_{||}^*} \cdot \nabla \Psi_{nl} \frac{\partial}{\partial p_{||}} \quad .$$

Defining the particle weight as $w_e = \delta f_e / f_e$, the weight equation can be derived by using Eq. (5):

$$\frac{dw_e}{dt} = Lw_e = -\left(1 - w_e\right)\frac{1}{f_{e0}}\left(\delta L_1 + \delta L_2\right)f_{e0}. \tag{6}$$

The parallel scalar potential $\delta A_{||}$ is solved from Ampere's law:

$$\left( \nabla_\perp^2 - \frac{\omega_{pe}^2}{c^2} - \frac{\omega_{pi}^2}{c^2} \right)\delta A_{||} = \frac{4\pi}{c}\left( en_{e0}U_{||ec} - Z_i n_{i0} U_{||ic} \right), \tag{7}$$

where $U_{||ec}$ and $U_{||ic}$ are the parallel canonical flow calculated from the kinetic markers:

$$U_{||\alpha c} = \frac{1}{n_{\alpha 0} m_\alpha} \int \mathbf{dv} p_{||} \delta f_\alpha , \tag{8}$$

where $\alpha = i, e$ and $\int \mathbf{dv} = \left(2\pi B_0 / m_\alpha^2\right)\int dp_{||} d\mu$. The appearances of the second and third terms on the LHS of Eq. (7) are due to the adiabatic parts associated with $\delta A_{||}$ from particle canonical momentum $p_{||}$. The last terms on both sides of Eq. (7) associated with ions are dropped in this paper for simplicity.

Integrating Eq. (5) in velocity space and keeping terms up to $O\left(\varepsilon^2\right)$ order, we can derive the electron continuity equation in a conservative form:

$$\frac{\partial \delta n_e}{\partial t} + \nabla \cdot \left[ n_{e0}\left( \delta u_{||e}\mathbf{b_0} + \mathbf{V_E} + u_{||e0}\frac{\boldsymbol{\delta B}}{B_0} \right) + \frac{1}{T_{e0}}\left( \delta P_{\perp e}\mathbf{V_g} + \delta P_{||e}\mathbf{V_c} \right) + \frac{n_{e0}\delta u_{||e}}{B_0}\boldsymbol{\delta B} + \delta n_e \mathbf{V_E} \right] = 0 , \tag{9}$$

where $\mathbf{V_E} = c\mathbf{b_0} \times \nabla \phi / B_0$ , $\mathbf{V_c} = \frac{cT_0}{q_e B_0}\mathbf{b_0} \times \left( \mathbf{b_0} \cdot \nabla \mathbf{b_0} \right)$ , and $\mathbf{V_g} = \frac{cT_0}{q_e B_0^2}\mathbf{b_0} \times \nabla B_0$ .

$u_{||e0} = -\frac{c}{4\pi e n_{e0}}\mathbf{b_0} \cdot \nabla \times \mathbf{B_0}$ is approximated as the equilibrium current, $\delta P_{||e} = \frac{1}{m_e}\int \mathbf{dv} p_{||}^2 \delta f_e$, $\delta P_{\perp e} = \int \mathbf{dv} \mu B_0 \delta f_e$ , and $\quad \delta \mathbf{B} = \nabla \times \left( \delta A_{\parallel} \mathbf{b_0} \right)$ is the magnetic perturbation.

$\delta u_{\parallel e} = U_{\parallel ec} - \dfrac{q_e \delta A_{\parallel}}{c m_e}$ represents the mechanical flow in Eq. (9). However, due to the fact that

$\left| U_{\parallel ec} \right| \approx \left| \dfrac{q_e \delta A_{\parallel}}{c m_e} \right| \gg \left| \delta u_{\parallel e} \right|$ in $\beta_e \gg m_e / m_i$ and $k_{\perp}^2 d_e^2 \ll 1$ regimes, it will bring severe

"cancellation problem" when we calculate $\delta u_{\parallel e}$ by using $\delta u_{\parallel e} = U_{\parallel ec} - \dfrac{q_e \delta A_{\parallel}}{c m_e}$ with particle noise. In order to overcome this problem, we calculate $\delta u_{\parallel e}$ by inverting the original Ampere's

law as:

$$n_{e0} e \delta u_{\parallel e} = \frac{c}{4\pi} \nabla_{\perp}^2 \delta A_{\parallel}, \tag{10}$$

where $\delta A_{\parallel}$ is solved from Eq. (7), and the ion current is dropped for simplicity. To the leading order terms up to $O\left( \varepsilon^2 \right)$, Eq. (9) integrated from drift kinetic equation in $p_{\parallel}$ formulation is the same with Eq. (21) of $v_{\parallel}$ formulation in Bao et al [33].

The Poisson's equation is [23]:

$$\frac{4\pi Z_i^2 n_{i0}}{T_{i0}} \left( \phi - \tilde{\phi} \right) = 4\pi \left( Z_i \delta n_i - e \delta n_e \right) \tag{11}$$

where $n_{i0}$ and $T_{i0}$ are the ion equilibrium density and temperature, respectively. $Z_i$ represents ion charge, and $\delta n_i = \int \mathbf{dv} \delta f_i$ is ion perturbed density. The terms on the left hand side of Eq. (11)

represents ion polarization density.

Eqs. (1-3) and (6-11) form a closed system together with ion gyrokinetic equation and represent the new conservative scheme, which can be applied to the nonlinear toroidal electromagnetic simulations with kinetic electrons. Note that by using the canonical momentum as an independence phase space coordinate, we avoid the well-known numerical issues of explicit time derivative of the vector potential to calculate the inductive parallel electric field. In this paper, we focus on verifying this new model for the linear shear Alfvén wave simulation in both shearless and sheared magnetic field. The application to the nonlinear and toroidal simulations will be reported in the future.

If we neglect ion finite Larmor radius (FLR) effects and parallel motion, then ion $\mathbf{E} \times \mathbf{B}$ drift cancels out with electron $\mathbf{E} \times \mathbf{B}$ drift, and ion only contributes to the polarization density (appears on the left hand side of Eq. (11)). The linear dispersion relation in the uniform plasmas can be

derived as:

$$\left(\frac{\omega^2}{k_{||}^2 V_A^2} - 1\right)\left[1 + \xi_e Z(\xi_e)\right] = k_\perp^2 \rho_s^2 , \tag{12}$$

where $\xi_e = \omega/\sqrt{2}k_{||}v_{the}$ , $\rho_s = C_s/\Omega_{ci}$ , $v_{the} = \sqrt{T_{e0}/m_e}$ , $C_s = \sqrt{T_{e0}/m_i}$ , and $Z(\xi_e)$ is the plasma dispersion function:

$$Z(\xi_e) = \frac{1}{\sqrt{\pi}} \int_{-\infty}^{+\infty} \frac{e^{-t^2}}{t - \xi_e} dt .$$

Thus, compared to the dispersion relation derived from Lin-Chen fluid-kinetic hybrid electron model [6], the ion acoustic wave is artificially removed in this paper for simplicity, but can easily be incorporated by retaining the kinetic ion contribution.

# III. Simulation results

## A. Shear-less magnetic field

To verify the new conservative scheme, we implement it in a 2D slab code to simulation kinetic shear Alfvén wave (KAW) in a uniform magnetic field pointing in the positive z-direction. The KAW wave vector is in the x-z plane. The electrons are treated using the drift kinetic equation and the ions as fixed background with only contribution to the polarization density. We first carry out simulations with $T_{e0} = 5.0 keV$ , $B = 1.5T$ , $n_{e0} = 1 \times 10^{13} cm^{-3}$ , $\beta_e = 0.9\%$ , $k_{||}/k_\perp = 0.01$ , and verify the wavelength $k\rho_s$ dependence of frequency as well as $\Delta x/d_e$ ( $\Delta x$ is the perpendicular grid size and $d_e$ is the electron skin depth) dependence. We change the wavelength in the range of $0.01 < k\rho_s < 0.48$ . The grid numbers per wavelength in the parallel and perpendicular directions are fixed as $N_z = 64$ and $N_x = 60$ , thus the grid size also changes with the wavelength. We use 20 markers per cell in all simulations of this paper. We give an initial density perturbation, and then let the perturbation evolve self-consistently. The KAW frequencies measured from simulations using electron continuity equation agree with theory well for both short and long wavelengths regimes. Most importantly, the perpendicular grid size can be much larger than the electron skin depth in the long wavelength regime as shown in Fig. 1. This shows that the perpendicular grid size is not restricted by the electron skin depth when we calculate the density perturbation from the electron continuity equation. For comparison, the simulation results deviate from the theoretical predication in the regime $k\rho_s \ll 1$ or $\Delta x/d_e \gg 1$ when we calculate the density perturbation from kinetic markers directly, which are shown by the black circles in Fig. 1.

In these cases without applying the electron continuity equation, the high frequency noise level greatly increases for the longer wavelength regime, which makes it difficult to measure the frequency when $k\rho_s < 0.05$.

Next, we apply an antenna with the theoretical frequency to excite the Alfvén waves in different wavelength regimes: $1/k\rho_s = 4.2$ ( $\Delta x/d_e = 1.2$ ), $1/k\rho_s = 20.8$ ( $\Delta x/d_e = 6.2$ ) and $1/k\rho_s = 41.5$ ( $\Delta x/d_e = 12.5$ ), respectively. The upper panels of Figs. 2(a)-(c) give the time histories of the electrostatic potential $\phi$ at the position with the maximal value from conventional scheme and conservative scheme. As the wavelength or perpendicular grid size increases, our simulation model does not suffer numerical noise or instability as shown by blue lines. However, the noise level of conventional $\delta f$ scheme increases with the perpendicular grid size as shown by red lines, even though the time step of the red lines is 10 times shorter than the blue lines in Fig. 2(c) for numerical stabilization. In conservative scheme simulations using the continuity equation, we plot in lower panels of Fig. 2(a)-(c) the perturbed densities calculated from the kinetic markers in order to compare with the results from the continuity equation. The discrepancy of the perturbed densities calculated from the kinetic markers and the continuity equation becomes larger as the wavelength or perpendicular grid size increases.

The reason of the inconsistency between the perturbed density and canonical flow calculated from the kinetic markers in the $\delta f$ method can be interpreted as follows. The assumption of the equilibrium distribution function $f_{e0}$ as the analytical Maxwellian $f_{Maxwellian}$ is not identical to the marker distribution $f_{Marker}$ in the simulations due to particle noise, and the error between these two distributions $\Delta f = f_{Marker} - f_{Maxwellian}$ accumulates in the simulation through Eq. (6) for the particle weight. Thus, after integrating the perturbed density and canonical flow from the kinetic markers using the particle weight, the electron continuity equation is not satisfied due to this error.

For shear Alfvén waves in the long wavelength limit, the parallel electric field $\delta E_{||}$ is nearly 0, which requires that the electrostatic electric field $\delta E_{||}^e = -\mathbf{b_0} \cdot \nabla\phi$ and the inductive electric field $\delta E_{||}^i = -(1/c)\partial\delta A_{||}/\partial t$ evolve consistently and cancel with each other. In conservative scheme, the consistency of $\delta E_{||}^e$ and $\delta E_{||}^i$ is enforced by using the continuity equation, and the total $\delta E_{||}$ is much smaller than its electrostatic or inductive components as shown in Fig. 3(a). When we calculated the perturbed density from the kinetic markers, the inconsistency of $\delta E_{||}^{e}$ and $\delta E_{||}^{i}$ gives rise to an unphysically large $\delta E_{||}$ as shown in Fig. 3(b). These observations prove our conclusion that the perturbed density and canonical flow calculated from the kinetic markers do not satisfy the continuity equation, and the corresponding electrostatic potential and vector potential do not evolve consistently in conventional $\delta f$ method, which causes the numerical instabilities in the electromagnetic simulation of the long wavelength modes in high $\beta_{e}$ plasmas. Furthermore, the theoretical study of the error effects on the dispersion relation and the parallel electric field are given in Sec. IV, which confirms that the inconsistency of the perturbed density and canonical flow in conventional $\delta f$ scheme can cause significant deviations of the parallel electric field from the exact solution.

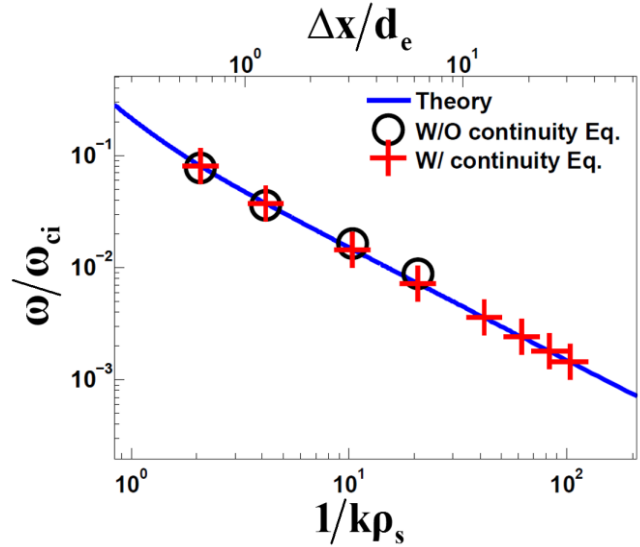

**Fig. 1.** Dependence of KAW frequency on wavelength (bottom) and perpendicular grid size (top). Red crosses represent results from new conservative scheme with continuity equation, and black circles represent results from conventional $\delta f$ scheme without continuity equation.

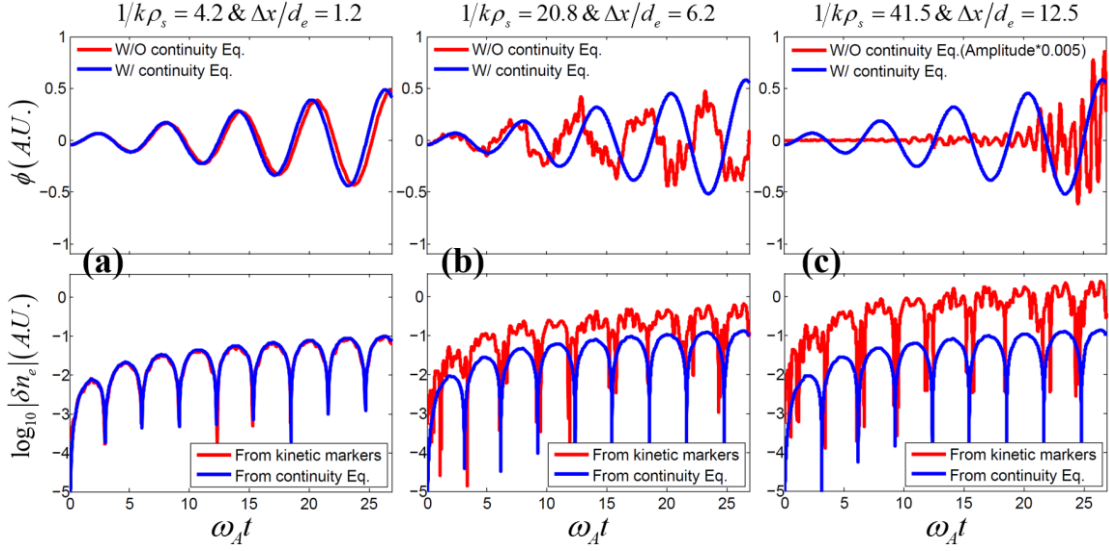

**Fig. 2.** Time histories of electrostatic potential (upper panels) and density perturbation (lower panels) for various wavelengths. In the upper panels, red lines are from conventional $\delta f$ scheme, and blue lines are from conservative scheme using continuity equation. In the lower panels, both red lines and blue lines represent density perturbation calculated from kinetic markers and from continuity equation in the same simulation using conservative scheme.

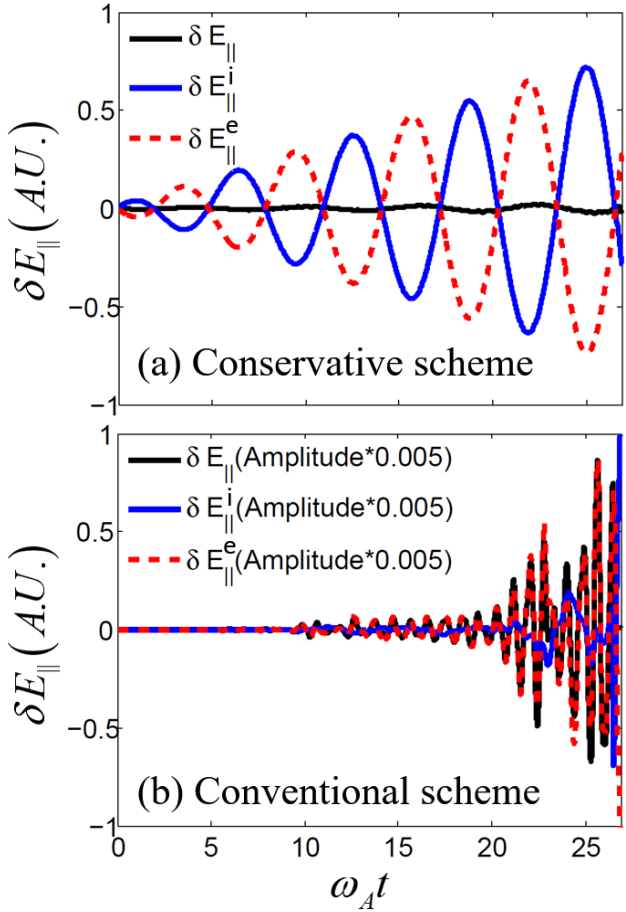

**Fig. 3.** The time histories of the parallel electric field $\delta E_{||}$ for $1/k\rho_s = 41.5$, (a) is from conservative scheme and (b) is from conventional $\delta f$ scheme.

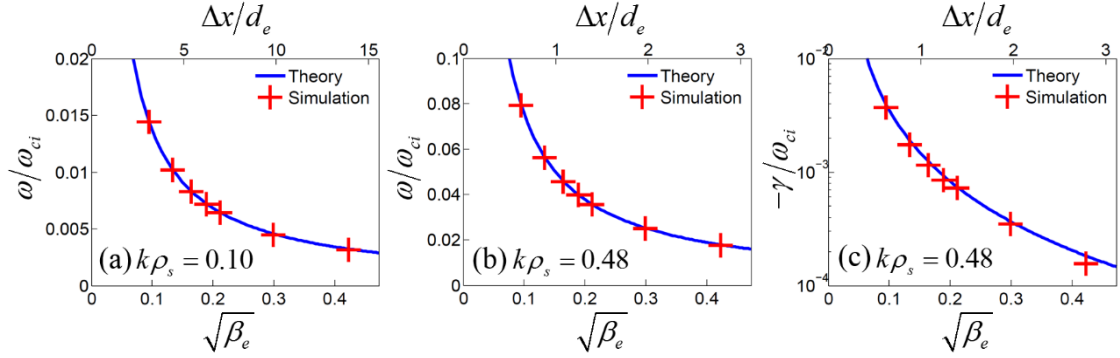

**Fig. 4.** Dependence of frequency on electron beta $\beta_e$ and perpendicular grid size $\Delta x/d_e$ for (a) $k\rho_s = 0.10$ and (b) $k\rho_s = 0.48$. (c) is dependence of damping rate on electron beta $\beta_e$ and perpendicular grid size $\Delta x/d_e$ for $k\rho_s = 0.48$. Red crosses represent simulation results from conservative scheme.

Secondly, we carry out simulations of shear Alfvén waves in different $\beta_e$ regimes. In the simulations, $k\rho_s = 0.10$ and $k\rho_s = 0.48$ are chosen as long and short wavelengths cases, respectively. We increase plasma density $n_{e0}$ from $1 \times 10^{13} cm^{-3}$ to $20.0 \times 10^{13} cm^{-3}$ for different $\beta_e$ values. Other parameters are the same with Fig.1. The grid number per parallel wavelength $N_z$ increases with $\beta_e$ for numerical convergence, e. g., $N_z = 128$ is used for $\beta_e = 18\%$ simulation, while the grid number per perpendicular wavelength $N_x = 60$ is enough for the convergence of all the cases in Fig. 4. Using conservative scheme, the frequencies and damping rates measured from the simulations agree well with the theory for $\beta_e \gg m_e/m_i$, and the corresponding perpendicular grid size does not need to resolve the electron skin depth in the very high $\beta_e$ cases. For the $k\rho_s = 0.10$ case, the damping rates of the shear Alfvén wave is too small to measure, which are absent in Fig. 4. Therefore, the accuracy of the dispersion relation of the shear Alfvén wave can be achieved by using the electron continuity equation in the high $\beta_e$ ($\beta_e \gg m_e/m_i$) and long wavelength regimes without resolving the electron skin depth.

## B. Sheared magnetic field

We now implement our model into the gyrokinetic toroidal code (GTC) [24] and carry out simulations of Alfvén wave in a cylinder plasma with magnetic shear. For comparisons with the kinetic simulations, we also suppress the kinetic effects in GTC to carry out fluid simulations of the Alfvén wave with the same parameters. In the fluid simulation, we apply the momentum equation by taking the first moment of the drift kinetic equation:

$$\frac{\partial U_{\parallel ec}}{\partial t} + \frac{q_e}{m_e} \mathbf{b_0} \bullet \nabla \phi + \frac{B_0}{m_e n_{e0}} \mathbf{b_0} \bullet \nabla \left( \frac{\delta P_{\parallel}}{B_0} \right) = 0 \tag{13}$$

where the pressure $\delta P_{\parallel} = \delta n_e T_{e0}$ using isothermal assumption. Eqs. (7), (9), (10), (11) and (13) form a closed system for fluid simulation with finite electron mass and temperature, which is equivalent to the fluid model with electron inertia proposed by D. Liu and L. Chen [25, 26].

In inhomogeneous plasmas, both $k_{\parallel} = \left[ nq(r) - m \right] / q(r) R$ and Alfvén speed $V_A$ varies spatially with the equilibrium magnetic field and density, which gives rise to the continuum damping of the Alfvén waves. In our simulation, the magnetic field is $B_0 = 2.0T$, equilibrium density and electron temperature are uniform with $n_{e0} = n_{i0} = 2.0 \times 10^{13} cm^{-3}$ and $T_{e0} = 5.0 keV$, the electron beta value is $\beta_e = 1.0\%$, the length of the cylinder is $l = 2\pi R_0$ ($R_0 = 1.0 m$), the minor radius is $a = 0.5m$ (our simulation domain is from $r_{min} = 0.39a$ to $r_{max} = 0.86a$, and $a/\rho_s = 139$), and the safety factor $q$ varies monotonously as $q = 0.76(r/a)^2 + 0.8$. The perpendicular grid size is $\Delta x/d_e = 2.0$ and the time step is $\Delta t = 0.04 R_0 / v_{the}$. We apply an initial density perturbation with the poloidal mode number $m = 3$ ($k_\theta \rho_s = 0.034$ at $r = 0.63a$, and initial $k_r \rho_s = 0.048$) and axial mode number $n = 2$. The time history of the electrostatic potential $\phi$ at the radial position $r = 0.63a$ is shown by Fig. 5(a). The Alfvén wave damps more quickly in the simulation with drift kinetic electrons than fluid electrons. As predicted, the time dependence of the mode amplitude due to the continuum damping is proportional to $1/t$ [27-29].

In order to verify the simulation with theory, we compare the $\phi$ amplitude evolution between fluid simulation and kinetic simulation in Fig. 5(b). It is found that the wave amplitude histories from both fluid (red line) and kinetic (blue line) simulations agree with the $1/t$ fitting (the black dashed line) at the initial phase, which verifies the Alfvén continuum damping. However, in the kinetic electron simulation, the electron Landau damping becomes the dominant damping mechanism later on, which is close to an exponential decay as shown by the black solid fitting line. For comparison, in the fluid simulation, the time history of $\phi$ agrees with the $1/t$ fitting, and the slight deviation in the later period is due to the finite electron mass and thermal pressure effects, which cause the propagation of Alfvén waves and affect the continuum damping. The radius-time plots of the absolute value of the electrostatic potential $\phi$ from kinetic and fluid simulations are shown by Figs. 5(c) and (d), respectively. Figure 5d shows that the continuum damping in the fluid simulation causes the increase of the radial wave vector, eventually leading to a singular radial structure (which is not physical). In contrast, the kinetic effects in the kinetic simulation results in the Landau damping and the formation of normal modes with discrete frequencies (which are more physical Landau eigen states) as shown in Fig. 5c. This indicates that the kinetic approach for the simulations of Alfvén waves is not just more physical, but also numerically superior than the fluid approach.

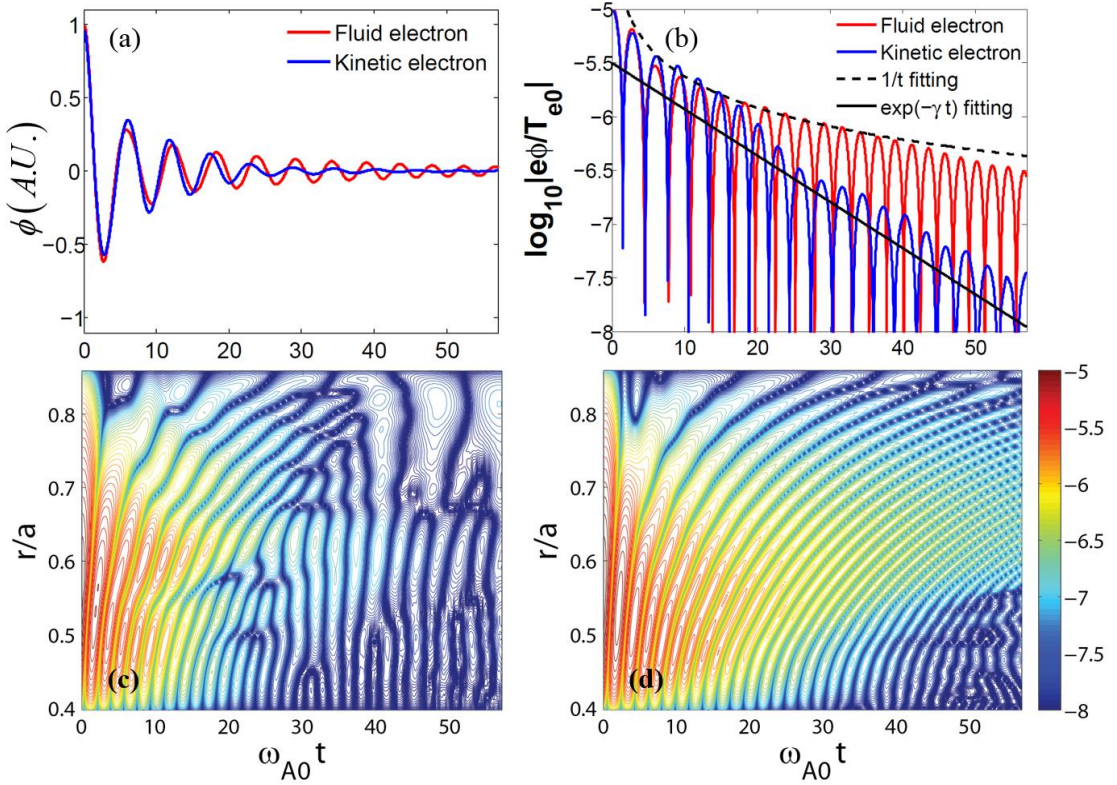

**Fig. 5.** (a) Time evolution of electrostatic potential $\phi$ at $r = 0.63a$, and (b) corresponding absolute value of electrostatic potential. (c) and (d) are radius-time plots of $\log_{10}\left(e|\phi|/T_{e0}\right)$ from kinetic and fluid simulations, respectively.

Furthermore, we carry out simulations of KAW using antenna excitation to verify the propagation in cylinder plasmas. The radial simulation domain is from $r_{\min} = 0.39a$ to $r_{\max} = a$.

The other parameters are the same as the continuum damping simulation earlier in this subsection. We excite the KAW with the poloidal mode number $m = 3$ ( $k_\theta \rho_s = 0.022$ at $r = a$ ) and axial mode number $n = 3$ at the edge by using an external antenna with local shear Alfvén frequency $\omega_{A0} = k_{\parallel 0} V_A$, where $k_{\parallel 0}$ is the parallel wave vector at $r = a$. The excited shear Alfvén wave can convert to kinetic Alfvén wave at the launching position since it is also the resonant layer where the relation $\omega_{A0} = k_{\parallel}(a) V_A$ is satisfied, then the KAW can propagate across the magnetic field lines[30, 31].

In Fig. 6, we present the radial electric field $\delta E_r = -\partial \phi / \partial r$ of Alfvén wave propagation from fluid and kinetic simulations with different perpendicular grid sizes. In the fluid simulation, the dispersion relation of the shear Alfvén wave with finite electron mass is

$$k_\perp^2 = \frac{1}{\rho_s^2} \left( 1 - \frac{\omega^2}{k_{\parallel}^2(r) V_A^2} \right) \Bigg/ \left( \frac{\omega^2}{k_{\parallel}^2(r) v_{vthe}^2} - 1 \right),$$

thus there exists the singular surface ( $k_\perp^2 \to +\infty$ ) where $\omega_{A0} = k_{\parallel}(r) v_{the}$ is satisfied for the Alfvén wave propagation with a fixed frequency. For comparison, the dispersion relation of the shear Alfvén wave with zero electron inertial term is

$$k_\perp^2 = \frac{1}{\rho_s^2} \left( \frac{\omega^2}{k_{\parallel}^2(r) V_A^2} - 1 \right),$$

and the singular surface for the shear Alfvén wave with $m = n = 3$ is

$q = 1$ surface where $k_{\parallel}(r) = 0$. In the simulation, we use the finite electron mass model. From the upper panels of Figs. 6(a)-(c), it can be seen that the perpendicular wave vector will increase as the Alfvén wave propagates toward the singular surface. The singular surface in the simulation deviates from the theoretical predictions, which is due to the fact that the spatial resolution is not enough to resolve the short radial wavelength. The shear Alfvén wave propagates to the theoretical singular surface only when the perpendicular grid size $\Delta x \to 0$. For comparison, there is no singularity in the drift kinetic electron simulation as shown by the lower panels of Figs. 6(a)-(c). The perpendicular grid size is less important, since the kinetic effects lead to the formation of a normal mode and thus prevent the increase of the perpendicular wave vector. This also shows the necessity of the kinetic approach for the simulations of Alfvén waves as illustrated in the earlier study of continuum damping in this subsection.

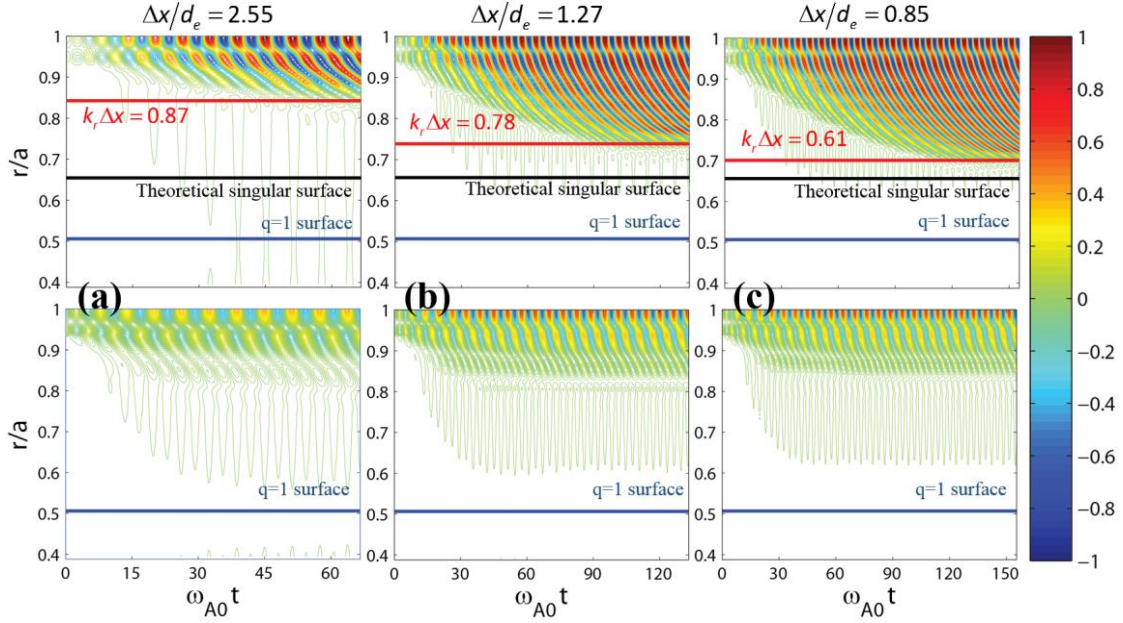

**Fig. 6**. Radius-time plots of radial electric field $\delta E_r\,(A.U.)$ of the shear Alfvén wave propagation using various perpendicular grid sizes. Upper panels are from fluid simulation and lower panels are from kinetic simulation.

# IV. Error estimations

## A. Error effects on the dispersion relation

We have demonstrated in Sec. III that the perpendicular grid size should resolve electron skin depth for conventional $\delta f$ scheme in order to control the error level, while there is no restriction of the perpendicular grid size in conservative scheme. In this part, we derive the dispersion relation of the shear Alfvén wave with particle noise effect for conventional $\delta f$ scheme and conservative scheme (continuity equation method), and explain why conservative scheme is superior to conventional $\delta f$ scheme theoretically.

Applying the Fourier transform: $\partial/\partial t = -i\omega t$, $\mathbf{b_0}\cdot\nabla = ik_{\|}$ to the linear version of Eq. (5), we can derive the perturbed electron distribution in the uniform plasmas:

$$\delta f = \frac{k_{\|}v_{\|}}{\omega - k_{\|}v_{\|}}\left(\phi - \frac{v_{\|}}{c}\delta A_{\|}\right)\frac{q_e}{T_{e0}}f_0, \tag{14}$$

where $v_{\|} = p_{\|}/m_e$. Electron $E \times B$ drift cancels with ion species and does not appear. The perturbed density and canonical flow can be derived by integrating Eq. (14) as:

$$\delta n = \frac{en_{e0}}{T_{e0}} \left( \phi - \frac{1}{c} \frac{\omega}{k_{||}} \delta A_{||} \right) \Big[ 1 + \xi_e Z(\xi_e) \Big] \tag{15}$$

and

$$U_{||c} = \left\{ -\frac{e}{cm_e} \delta A_{||} + \frac{e}{T_{e0}} \frac{\omega}{k_{||}} \left( \phi - \frac{1}{c} \frac{\omega}{k_{||}} \delta A_{||} \right) \Big[ 1 + \xi_e Z(\xi_e) \Big] \right\}. \tag{16}$$

In conventional $\delta f$ scheme, both electron density and canonical flow perturbations are calculated from the kinetic markers, thus we multiply the exact solutions Eqs. (15) and (16) by error coefficients to represent the values measured from the simulation with particle noise. The perturbed density and canonical flow calculated from kinetic markers with particle noise effect can be expressed as [32]:

$$\delta \tilde{n} = \frac{en_{e0}}{T_{e0}} \left( \phi - \frac{1}{c} \frac{\omega}{k_{||}} \delta A_{||} \right) \Big[ 1 + \xi_e Z(\xi_e) \Big] (1 + \varepsilon_n) \tag{17}$$

and

$$\tilde{U}_{||c} = \left\{ -\frac{e}{cm_e} \delta A_{||} + \frac{e}{T_{e0}} \frac{\omega}{k_{||}} \left( \phi - \frac{1}{c} \frac{\omega}{k_{||}} \delta A_{||} \right) \Big[ 1 + \xi_e Z(\xi_e) \Big] \right\} (1 + \varepsilon_u), \tag{18}$$

where $\varepsilon_n$ and $\varepsilon_u$ represent the particle noise errors induced from the kinetic markers to the perturbed density and the perturbed canonical flow, respectively, Eqs. (17) and (18) become the exact solutions when $\varepsilon_n = \varepsilon_u = 0$.

Applying the Fourier transform $\nabla = i\mathbf{k}_\perp$ to Eqs. (7) and (11) and considering Eqs. (17) and (18), we can derive the dispersion relation of shear Alfvén wave with particle noise error:

$$\left( \frac{\omega^2}{k_{||}^2 V_A^2} - 1 \right) \Big[ 1 + \xi_e Z(\xi_e) \Big] = k_\perp^2 \rho_s^2 \frac{1}{1 + \varepsilon_u} - \frac{\beta_e}{2} \frac{m_i}{m_e} \frac{\varepsilon_u}{1 + \varepsilon_u} \\ - \frac{1}{k_\perp^2 d_e^2} \Big[ 1 + \xi_e Z(\xi_e) \Big] \frac{\varepsilon_u(1 + \varepsilon_n)}{1 + \varepsilon_u} + \Big[ 1 + \xi_e Z(\xi_e) \Big] \frac{\varepsilon_n - \varepsilon_u}{1 + \varepsilon_u} \tag{19}$$

Then we can expand Eq. (19) using $|\varepsilon_n| \ll 1$ and $|\varepsilon_u| \ll 1$ and only keep the first order error terms:

$$\left( \frac{\omega^2}{k_{||}^2 V_A^2} - 1 \right) \Big[ 1 + \xi_e Z(\xi_e) \Big] = k_\perp^2 \rho_s^2 (1 - \varepsilon_u) - \frac{\beta_e}{2} \frac{m_i}{m_e} \varepsilon_u \\ - \frac{1}{k_\perp^2 d_e^2} \Big[ 1 + \xi_e Z(\xi_e) \Big] \varepsilon_u + \Big[ 1 + \xi_e Z(\xi_e) \Big] (\varepsilon_n - \varepsilon_u) \tag{20}$$

From Eq. (20), it is seen that the particle noise error is amplified with increasing $\beta_e$ and the perpendicular wavelength through the second and third terms on the RHS, respectively, which explains the numerical difficulty of the electromagnetic simulation of the long wavelength modes with $k_\perp^2 d_e^2 \ll 1$ in high $\beta_e$ plasmas. For example, the numerically high frequency modes dominate in the simulation of long wavelength ($k_\perp^2 d_e^2 \ll 1$) shear Alfvén waves by using conventional $\delta f$ scheme as shown by Fig. 2, which are caused by the third error term on the RHS of Eq. (20).

In conservative scheme, we use the kinetic markers to calculate the perturbed canonical flow as Eq. (18). The parallel vector potential with error effect can be calculated by using Eqs. (7) and (18):

$$\delta A_{||} = \frac{c\dfrac{\omega}{k_{||}}\left[1 + \xi_e Z(\xi_e)\right](1+\varepsilon_u)}{-c^2 k_\perp^2 \lambda_D^2 + v_{the}^2 \varepsilon_u + \dfrac{\omega^2}{k_{||}^2}\left[1 + \xi_e Z(\xi_e)\right](1+\varepsilon_u)}\phi .$$

(21)

From Eqs. (9), (10) and (21), the perturbed density with particle noise error in the linear and uniform plasmas is:

$$\delta n = \frac{-\dfrac{c^2 k_\perp^2}{4\pi e}\left[1 + \xi_e Z(\xi_e)\right](1+\varepsilon_u)}{-c^2 k_\perp^2 \lambda_D^2 + v_{the}^2 \varepsilon_u + \dfrac{\omega^2}{k_{||}^2}\left[1 + \xi_e Z(\xi_e)\right](1+\varepsilon_u)}\phi .$$

(22)

From Eqs. (11) and (22), the dispersion relation of the shear Alfvén wave with particle noise error in the conservative scheme is:

$$\left(\frac{\omega^2}{k_{||}^2 V_A^2} - 1\right)\left[1 + \xi_e Z(\xi_e)\right] = k_\perp^2 \rho_s^2 \frac{1}{1+\varepsilon_u} - \frac{\beta_e}{2}\frac{m_i}{m_e}\frac{\varepsilon_u}{1+\varepsilon_u} .$$

(23)

Keeping the first order terms with respect to particle noise error, Eq. (23) can be written as:

$$\left(\frac{\omega^2}{k_{||}^2 V_A^2} - 1\right)\left[1 + \xi_e Z(\xi_e)\right] = k_\perp^2 \rho_s^2 (1-\varepsilon_u) - \frac{\beta_e}{2}\frac{m_i}{m_e}\varepsilon_u .$$

(24)

Compared to Eq. (20), there is no particle noise error term related to the perpendicular wavelength in Eq. (24). It is because that the continuity equation guarantees the consistency between $\varepsilon_n$ and $\varepsilon_u$ in conservative scheme, and the last two terms on the RHS of Eq. (19) cancel with each other.

Thus conservative scheme is more robust than conventional $\delta f$ scheme for the simulations of long wavelength electromagnetic modes. For example, conservative scheme can simulate shear

Alfvén waves in the long wavelength regimes accurately while conventional $\delta f$ scheme fails as shown by Fig. 2. From Eqs. (20) and (24), it is also noticed that finite $\beta_e$ can amplify error terms for both schemes. However, we can simulate the shear Alfvén waves with sufficient accuracy in the very high $\beta_e$ regime by using conservative scheme as shown by Fig. 4, which indicates that finite $\beta_e$ effects are not as severe as long wavelength ($k_\perp^2 d_e^2 \approx 0$) effects on the numerical properties of electromagnetic simulations.

## B. Error effects on the parallel electric field $\delta E_{||}$

As we have illustrated before, the perturbed density and canonical flow calculated from the kinetic markers do not satisfy the electron continuity equation, which leads to the inconsistency of the corresponding electrostatic potential and vector potential in conventional perturbative ($\delta f$) simulation. However, the parallel electric field $\delta E_{||} \approx 0$ in the long wavelength (ideal MHD) limit, which requires that the electrostatic electric field $\delta E_{||}^e = -\mathbf{b_0} \cdot \nabla \phi$ and the inductive electric field $\delta E_{||}^i = (1/c) \partial \delta A_{||} / \partial t$ evolve consistently and their major parts can cancel with each other in the simulation. Thus, any inconsistency of $\delta E_{||}^e$ and $\delta E_{||}^i$ can lead to the unphysically large $\delta E_{||}$, and cause the numerical instability in conventional perturbative ($\delta f$) scheme. In this part, the error effects on the parallel electric field $\delta E_{||}$ are discussed for both conventional $\delta f$ scheme and conservative scheme.

In conventional $\delta f$ scheme, applying the Fourier transform $\nabla = i\mathbf{k}_\perp$ to Eqs. (7) and (11), and considering Eqs. (17) and (18), the electrostatic potential and vector potential can be represented as:

$$\phi = -\frac{4\pi e n_{e0} V_A^2}{c^2 k_\perp^2} \frac{k_{||}}{\omega} \left( \frac{1 + \varepsilon_n}{1 + \varepsilon_u} - \frac{1 + \varepsilon_n}{1 + k_\perp^2 d_e^2} \right) \tilde{U}_{||\kappa}, \tag{25}$$

$$\delta A_{||} = -\frac{d_e^2}{1 + k_\perp^2 d_e^2} \frac{4\pi e n_{e0}}{c} \tilde{U}_{||\kappa}. \tag{26}$$

Thus, the parallel electric field $\delta E_{||}$ in conventional $\delta f$ scheme is:

$$\delta E_{||} = i \frac{m_e}{e\omega} \frac{k_{||}^2 V_A^2}{1+k_\perp^2 d_e^2} \left( \frac{\omega^2}{k_{||}^2 V_A^2} - 1 \right) \tilde{U}_{||\kappa} + i \frac{m_e}{e\omega} \frac{k_{||}^2 V_A^2}{k_\perp^2 d_e^2} \left( \frac{1+\varepsilon_n+k_\perp^2 d_e^2}{1+k_\perp^2 d_e^2} - \frac{1+\varepsilon_n}{1+\varepsilon_u} \right) \tilde{U}_{||\kappa} . \qquad (27)$$

By keeping the first order error terms based on $|\varepsilon_n| \ll 1$ and $|\varepsilon_u| \ll 1$, Eq. (27) can be simplified as:

$$\delta E_{||} = i \frac{m_e}{e\omega} \frac{k_{||}^2 V_A^2}{1+k_\perp^2 d_e^2} \left( \frac{\omega^2}{k_{||}^2 V_A^2} - 1 \right) \tilde{U}_{||\kappa} + i \frac{m_e}{e\omega} \frac{k_{||}^2 V_A^2}{k_\perp^2 d_e^2} \left( \varepsilon_u - \varepsilon_n \frac{k_\perp^2 d_e^2}{1+k_\perp^2 d_e^2} \right) \tilde{U}_{||\kappa} . \qquad (28)$$

The last term on the RHS of Eq. (28) is due to inconsistency between $\varepsilon_n$ and $\varepsilon_u$, which will be amplified when $k_\perp^2 d_e^2$ is small.

In conservative scheme, the perturbed density is calculated from Eq. (9) instead of Eq. (17). The electrostatic potential can be derived by using Eqs. (9), (10), (11) and (26) as:

$$\phi = -\frac{4\pi e n_{e0} V_A^2}{c^2 k_\perp^2} \frac{k_{||}}{\omega} \left( \frac{k_\perp^2 d_e^2}{1+k_\perp^2 d_e^2} \right) \tilde{U}_{||\kappa} . \qquad (29)$$

By using Eqs. (26) and (29), the parallel electric field $\delta E_{||}$ in conservative scheme is:

$$\delta E_{||} = i \frac{m_e}{e\omega} \frac{k_{||}^2 V_A^2}{1+k_\perp^2 d_e^2} \left( \frac{\omega^2}{k_{||}^2 V_A^2} - 1 \right) \tilde{U}_{||\kappa} . \qquad (30)$$

Comparing Eqs. (28) and (30) in the long wavelength limit with $k_\perp^2 d_e^2 \ll 1$, it is seen that $\delta E_{||} \approx 0$ with $\omega^2 \approx k_{||}^2 V_A^2$ for conservative scheme, while $\delta E_{||} \propto 1/k_\perp^2 d_e^2$ with $\omega^2 \approx k_{||}^2 V_A^2$ for conventional $\delta f$ scheme due to the inconsistency between the perturbed density and canonical flow. Thus, conservative scheme can guarantee $\delta E_{||} \approx 0$ in the simulation as shown by Fig.3 (a), while conventional $\delta f$ scheme causes the unphysically large amplitude of $\delta E_{||}$ in the simulation as shown by Fig. 3(b), which is due to the last error term on the RHS of Eq. (28).

# V.    Conclusions

In this work, we present an innovative conservative scheme to solve the numerical difficulty of the electromagnetic gyrokinetic particle simulation with kinetic electrons. In conventionally electromagnetic perturbative $(\delta f)$ simulation with drift kinetic electrons of $p_{||}$ formulation, we found that perturbed density and perturbed canonical flow calculated from the kinetic markers do not satisfy the continuity equation due to the particle noise, which causes the inconsistency between electrostatic potential and parallel vector potential and leads to an unphysically large parallel electric field in the long wavelength MHD regime. In conservative scheme, we use the electron continuity equation to time advance the electron density perturbation using the perturbed mechanical flow calculated from the parallel vector potential, and the parallel vector potential is solved by using the canonical flow from the kinetic markers, which guarantees the consistency between electrostatic and parallel vector potentials. Furthermore, conservative scheme helps to relax the perpendicular grid size without resolving the electron skin depth when the wavelength is longer than the electron skin depth. However, in conventional electromagnetic $\delta f$ scheme, perpendicular grid size needs to resolve the electron skin depth even for the long wavelength MHD wave simulation. The simulations of shear Alfvén wave dispersion relation and continuum damping are well benchmarked with theory by using conservative scheme. In conservative scheme simulation of long wavelength shear Alfvén wave, the electrostatic component and inductive component of the electric field cancel with each other as expected. Finally, comparison of Alfvén wave propagation results between gyrokinetic particle simulation and fluid simulation shows that gyrokinetic particle simulation is superior to the fluid simulation by physically avoiding the singular mode structures through electron Landau damping. Building on the linear application of conservative scheme to shear Alfvén wave simulation, the nonlinear simulation results will be reported in the future work. Meanwhile, the application of conservative scheme on drift kinetic electron model with $v_{||}$ formulation combining with split weight scheme has been reported in another work [33].

# Acknowledgements

We would like to thank W. W. Lee, L. Chen, W. M. Tang, Y. Chen, D. Liu, I. Holod, L. Shi, C. Lau, Y. Xiao and W. L. Zhang for useful discussions. This work was supported by China National Magnetic Confinement Fusion Science Program (Grant No. 2013GB111000), US Department of Energy (DOE) SciDAC GSEP Program and China Scholarship Council (Grant No. 201306010032). This work used resources of the Oak Ridge Leadership Computing Facility at Oak Ridge National Laboratory (DOE Contract No. DE-AC05-00OR22725) and the National Energy Research Scientific Computing Center (DOE Contract No. DE-AC02-05CH11231).